\documentclass[12pt,preprint]{aastex}  

\usepackage{epsfig}
\usepackage{natbib}
\usepackage{slashbox}

\shorttitle{Energy Buildup for Giant Flares}

\shortauthors{C. Yu}

\begin{document}

\title{Magnetic Energy Buildup for Relativistic Magnetar Giant Flares}


\author{Cong ~Yu\altaffilmark{1,2}}
\altaffiltext{1}{National Astronomical Observatories/Yunnan
Astronomical Observatory, Chinese Academy of Sciences, Kunming,
650011, China; {\tt cyu@ynao.ac.cn}}

\altaffiltext{2}{Key Laboratory for the Structure and Evolution of
Celestial Objects, Chinese Academy of Sciences, Kunming, 650011,
China}

\date{$\qquad\qquad\qquad\qquad\qquad \qquad\qquad$ Apr. 13 2011}
\begin{abstract}
Motivated by coronal mass ejection studies, we construct general
relativistic models of a magnetar magnetosphere endowed with
strong magnetic fields. The equilibrium states of the stationary,
axisymmetric magnetic fields in the magnetar magnetosphere are
obtained as solutions of the Grad-Shafranov equation in a
Schwarzschild spacetime. To understand the magnetic energy buildup
in the magnetar magnetosphere, a generalized magnetic virial
theorem
in the Schwarzschild metric is newly derived. We carefully address
the question whether the magnetar magnetospheric magnetic field
can build up sufficient magnetic energy to account for the work
required to open up the magnetic field during magnetar giant
flares.
We point out the importance of the Aly-Sturrock constraint, which
has been widely studied in solar corona mass ejections, as a
reference state in understanding magnetar energy storage
processes.
We examine how the magnetic field can possess enough energy to
overcome the Aly-Sturrock energy constraint and open up. In
particular, general relativistic (GR) effects on the Aly-Sturrock
energy constraint in the Schwarzschild spacetime are carefully
investigated.
It is found that, for magnetar outbursts, the Aly-Sturrock
constraint is more stringent, i.e., the Aly-Sturrock energy
threshold is enhanced due to the GR effects. In addition, neutron
stars with greater mass have a higher Aly-Sturrock energy
threshold and are more difficult to erupt. This indicates that
magnetars are probably not neutron stars with extreme mass. For a
typical neutron star with mass of $1-2 M_{\odot}$,
we further explore the effects of cross-field current effects,
caused by the mass loading, on the possibility of stored magnetic
field energy exceeding the Aly-Sturrock threshold.

\end{abstract}



\section{INTRODUCTION}

After the discovery of soft gamma repeaters and anomalous X-ray
pulsars (Mazets et al. 1979; Mereghetti \& Stella 1995), magnetar
models of these sources are proposed to explain the relevant
phenomena (Duncan \& Thompson 1992; Thompson, Lyutikov \& Kulkarni
2002). Magnetars are believed to be neutron stars with strong
magnetic field, $\sim 10^{14} - 10^{15}$G (Duncan \& Thompson
1992). The magnetar outbursts, such as giant flares, occur with
huge release of magnetic energy $\sim 10^{44} - 10^{46}$ ergs. The
energy for magnetar outbursts is widely accepted to be supplied by
the star's magnetic field. However the physical process by which
the energy is stored and released is one of the great puzzles in
high energy astrophysics. Two possibilities exist for the location
where the magnetic energy is stored prior to an eruption: in the
magnetar crust or in the magnetosphere. For the former
possibility, a giant flare may be caused by a sudden untwisting of
the magnetar interior magnetic field (Thompson \& Duncan 2001).
Subsequently, a sudden
and brittle fracture of the crust leads to the giant flare.
In this crust scenario, the energy stored in the external twist is
limited by the tensile strength of the crust. Alternatively, based
on the short timescale of the giant flare rise time, $\sim
0.25\mathrm{ms}$ (Palmer et al. 2005), the second possibility $-$
the magnetospheric storage model, was proposed by Lyutikov (2006).
The energy released during an eruption is stored slowly (on a
longer timescale than the timescale of giant flare) in the
magnetar magnetosphere prior to the eruption.
An abrupt reconfiguration and dissipation of the magnetic field
due to a loss of confinement (Flyer et al. 2004) or a dynamical
instability (Lyutikov 2003; Komissarov et al. 2007)
produces the giant flare. 
This mechanism has the feature that the energy stored in the
external twist may not be limited by the tensile strength of the
crust, but instead by the total external magnetic field energy.

The magnetospheric storage model of magnetar giant flare shares
similar magnetic energy buildup process to solar eruptions, such
as coronal mass ejections (CMEs). In this model, the energy
released during an eruption is stored in the magnetospheric
magnetic field before the eruption. Large-scale eruptive CMEs
often give rise to the opening up of magnetic field lines that
were originally closed. The processes of magnetic fields opening
up have been extensively investigated in the CME studies (Barnes
\& Sturrock 1972; Aly 1984; Mikic \& Linker 1994). It is
physically reasonable to assume that the preeruption closed state
must possess more magnetic energy than the posteruption open
state. As will be discussed in detail below, requiring the
magnetic field to open imposes an extreme energy constraint on
theories for CMEs. This energy requirement on solar CMEs has been
under extensive theoretical studies in the past decades (Aly 1984;
Sturrock 1991; Wolfson \& Dlamini 1997; Zhang \& Low 2005). The
energy storage processes take place quasi-statically on a long
timescale. When the magnetic field reaches a threshold, due to the
instability or loss of confinement, the field erupts suddenly on a
much shorter dynamical timescale. Analogous processes of magnetic
field opening up are believed to occur in magnetar giant flares
(Woods et al. 2001; Thompson et al. 2002; Beloborodov 2009). All
these features of the storage model are in good agreement with the
observations of magnetar giant flares (Lyutikov 2006).



The similarity between solar eruptions and magnetar giant flares
(Lyutikov 2003) motivates this study on the energy buildup process
in the magnetar magnetosphere. We note that there are important
differences between solar eruptions and magnetar outbursts. For
situations in the magnetar magnetosphere (Beloborodov \& Thompson
2007), the location where the magnetic energy buildup occurs is
quite near the neutron star surface ($\sim 1-2 R_{\mathrm{NS}}$).
General relativistic (GR) effects near the neutron star surface
are important (Ciolfi et al. 2009).
General relativistic effects are currently, however, not taken
into account in relevant energy storage processes. In this work we
will investigate these processes with GR spacetime curvature
effects considered. More specifically, we will ignore effects of
magnetar rotation since they are slow rotators and describe the
background geometry of the magnetar magnetosphere with the
Schwarzschild metric.

The virial theorem is a helpful tool for us to understand the
energy properties in the magnetar magnetosphere. The flat
spacetime magnetic virial theorem (Chandrasekhar 1961) has been
extensively exploited in astrophysical researches (Aly 1984; Zhang
\& Low 2005). Attempts to get the GR virial theorem have been made
by Chandrasekhar (1967), but he just considered a hydrostatic
system, with the effects of magnetic fields completely ignored. In
this study we establish the magnetic virial theorem in the
Schwarzschild metric, which helps us to better understand GR
effects on the physical behaviors of the magnetic energy buildup.

For the magnetospheric storage model, an important question for
the giant flare energetics is: Can the magnetospheric magnetic
field store enough energy before an eruption?
For the magnetic energy alone to power a magnetar giant flare, the
energy must be sufficient to open up the magnetic field. However,
for the nearly force-free magnetic field exterior to a sphere, a
well-known result by Aly (1984, 1991) and Sturrock (1991) suggests
that the energy of a fully open field is the upper limit on the
energies of all the force-free fields in simple
geometries\footnote{Here simple geometries mean that the two ends
of all field lines are anchored onto the neutron star surface.}.
Thus the transition from a closed field configuration to an open
one (which is actually required for a realistic eruption) is not
energetically favored. 
Due to this Aly-Sturrock constraint,
the initial magnetic field before eruption must have energy in
excess of the threshold set by the Aly-Sturrock energy constraint.
This Aly-Sturrock constraint is widely discussed in the solar CMEs
studies. But its implications for magnetars are only briefly
mentioned in Lyutkov (2006). Furthermore, GR effects on this
important Aly-Sturrock constraint have not been considered in
prior works.
One purpose of this work is to clarify how GR effects influence
the Aly-Sturrock constraint.

The Aly-Sturrock constraint constitutes a bottleneck for the
storage model of magnetar giant flares. There are a number of
ways, however, to avoid the Aly-Sturrock
constraint. 
A deviation from a perfectly force-free initial state might make a
difference. In this scenario, it is expected that the cross-field
electric currents are viable source of energy for the eruption.
Detailed calculations about solar CMEs by Low \& Smith (1993)
suggested that a non-force-free magnetic field with cross field
currents due to the mass loading of plasma can store more energy
than the Aly-Sturrock field. Such mass loading effects are further
discussed by Wolfson \& Dlamini (1997) and Zhang \& Low (2004).
The mass loading of plasma in a non-force-free magnetic field acts
like a rigid wall to confine the magnetic field, in other words,
it would act as a lid that allows the magnetic energy to increase
above the limit, and when the lid is suddenly removed, the field
springs outward (Fan \& Low 2003). By analogy, it is possible that
in the magnetar magnetosphere, the mass loading plays the same
role to compress the magnetic field. Consequently, the magnetic
field can store magnetic energy above the Aly-Sturrock constraint.
But no theoretical calculations were performed to corroborate this
idea. In this work we will provide such a demonstration.
Another possibility for the magnetic energy to exceed the
Aly-Sturrock constraint is the formation of detached field lines
from the magnetar surface (magnetic bubble or magnetic flux rope,
e.g., Low \& Smith 1993; Flyer et al. 2004), which will be further
discussed in Yu et al. (2011, in prep).

This paper is organized as follows: in \S 2 we introduce the
generalized magnetic virial theorem in the Schwarzschild metric.
In \S 3 we will discuss how the the Aly-Sturrock field energy is
affected by general relativistic effects. We will explore the
cross-field effects caused by the mass loading on the magnetic
energy storage in \S 4. Conclusions and discussions are given in
\S 5.

%

\section{Generalized Virial Theorem in Schwarzschild Spacetime}
The virial theorem is of vital importance for understanding the
magnetic energy storage in the magnetar magnetosphere. In the flat
spacetime, it was proposed by Chandrasekhar (1961) and has been
used widely in solar physics researches (e.g., Low \& Smith 1993).
We focus in this paper on the physical behavior near the magnetar
surface, GR effects should be incorporated.
Because observed magnetars have a very slow rotation
rate, we ignore the rotation effects and adopt the Schwarzschild
metric as the background spacetime. In this section we establish
the virial theorem in the Schwarzschild metric including effects
of magnetic fields. We consider a steady state magnetosphere
around magnetars. The metric $g_{\mu\nu}$ of Schwarzschild
geometry reads (Misner, Thorne \& Wheeler 1973)
\begin{equation}
ds^{2} = g_{\mu\nu}dx^{\mu}dx^{\nu} = - \alpha^2 dt^2 +
\alpha^{-2} dr^2 + r^2 d\theta^2 + r^2\sin^{2}\theta d\phi^2 \ .
\end{equation}
The factor of $\alpha$ is defined as
\begin{equation}\label{alphafactor}
\alpha(r) = \sqrt{1-\frac{2 r_g}{r}} \ ,
\end{equation}
where $r_g  = G \mathcal{M}_{\mathrm{ns}}/c^2$ is the
gravitational radius, $G$ is the gravitational constant,
$\cal{M}_{\mathrm{ns}}$ is the mass of the neutron star, and $c$
is the speed of light.

A plasma containing only a perfect fluid and an electromagnetic
field, is described by the energy-momentum tensor (Anile 1989)
\begin{equation}\label{energymomentum}
T^{\mu\nu} = T^{\mu\nu}_{\mathrm{fluid}} +
T^{\mu\nu}_{\mathrm{EM}} =  \left( p+\rho + b^2 \right) u^{\mu}
u^{\nu} + \left( p + \frac{b^2}{2} \right) g^{\mu \nu} - b^{\mu}
b^{\nu} \ ,
\end{equation}
where $p$ is the isotropic pressure, $\rho = \rho_0 +
\frac{p}{(\gamma-1)}$ is the energy density (including that due to
the rest mass $\rho_0$) and $b^2 = b_{\mu} b^{\mu}$. A polytropic
equation of state is adopted and we take $\gamma = 4/3$ throughout
this paper. Here the Einstein summation rule is assumed and Greek
letters take on the values $t$, $r$, $\theta$, and $\phi$. The
magnetic field 4-vector is
\begin{equation}
b^{\mu} =^*F^{\mu\nu} u_{\nu} \ ,
\end{equation}
where $^*F^{\mu\nu}$ is the Maxwell tensor and $u_{\nu}$ is the
four velocity of the comoving observer (Anton et al. 2006). The
plasma is assumed to be in magnetostatic equilibrium, thus the
four velocity is $ u^{\mu} = \left((-g_{tt})^{-1/2},0,0,0
\right)$. Under such circumstances, the condition
\begin{equation}
\nabla_{\nu} T^{\mu\nu} = 0 \ ,
\end{equation}
reduces to
\begin{equation}\label{derivevirial}
g^{\mu\nu}\frac{\partial \left( p + \frac{b^2}{2} \right)}{\partial x^{\nu}} + %
\frac{(p+\rho+b^2)}{2}g^{\mu\nu}\frac{\partial \ln(-g_{tt})}{\partial x^{\nu}} + %
\frac{1}{\sqrt{-g}}\frac{\partial}{\partial x^{\nu}}\left( \sqrt{-g} \ b^{\mu}b^{\nu}\right) %
+\Gamma^{\mu}_{\lambda\sigma}b^{\lambda}b^{\sigma} = 0 \ , %
\end{equation}
where $g$ is the determinant of the metric $g_{\mu\nu}$ and the
explicit expressions of the connection coefficients
$\Gamma^{\mu}_{\lambda\sigma}$ (Weinberg 1972) are given in
Appendix A.

A re-arrangement of various terms of the above equation using
Gauss theorem leads to the following generalized virial theorem in
the Schwarzschild spacetime (Details are given in Appendix A),
\[
E + (3\gamma - 4) U = %
\int_{\partial \mathrm{V}} \alpha^2\left(\frac{ B^2}{2}+p\right) (\mathbf{r\cdot} d \mathbf{S}) %
\]
\begin{equation}\label{virial}
- \int_{\partial \mathrm{V}} \alpha^2 (\mathbf{B \cdot r}) (\mathbf{B\cdot}d \mathbf{S}) %
+ \int_{\mathrm{V}} \frac{\left( 1 - \alpha^2 \right)}{2} %
\left( B_r^2 + B^2  + \frac{5\gamma - 4}{\gamma -1} p 
\right) dV 
\ .
\end{equation}
In this equation, $\mathbf{r}$ is the position vector. Here
$d\mathbf{S}$ is a surface area element directed outwards and $d
V$ is a volume element, both measured by a locally inertial
observer. The factor of $\alpha$ is given in equation
(\ref{alphafactor}). Note that the total energy $E$ is the sum of
the magnetic, internal and gravitational potential energy, namely
\begin{equation}
E = M + U + W \ ,
\end{equation}
where
\begin{equation}\label{Mdefinition}
M = \int \frac{B^2}{2} \ dV \ ,
\end{equation}
\begin{equation}
U = \int\frac{p}{\gamma - 1} \ dV \ ,
\end{equation}
\begin{equation}
W = - \int\frac{\rho_0 G \mathcal{M}_{\mathrm{ns}}}{r} \ dV \ ,
\end{equation}
are the magnetic, internal and gravitational potential energy,
respectively.
Here we have absorbed a $4\pi$ factor into the definition of the
magnetic fields throughout this paper. In the above equations, the
magnetic field $\mathbf{B}$ in the ``ordinary" orthogonal basis
(defined in Section 4) is used. The relation between $\mathbf{B}$
and the magnetic field 4-vector $b^{\mu}$ is given explicitly in
Appendix A. Note that $B_r$ is the radial component of
$\mathbf{B}$ and $B^2$ = $B_r^2 + B_{\theta}^2 + B_{\phi}^2$.
Throughout this paper, we mainly work with the magnetic field
$\mathbf{B}$. This choice is made mainly for the convenience of
comparison between the results in the curved spacetime and the
flat spacetime.

The last integral on the right hand side in equation
(\ref{virial}) appears owing to general relativistic effects.
This term disappears when taking the flat spacetime limit, i.e.,
$\alpha^2$ $\rightarrow$ 1. Note also that this equation becomes
the usual virial theorem in the flat spacetime as $\alpha^2$
$\rightarrow$ 1 (Chandrasekhar 1961). In
particular\footnote{Although we will be treating non-force-free
magnetosphere in which cross-field effects (caused by mass
loading) are important, the discussion here is restricted to
magnetically dominated force-free fields. The relevance will
become clear as we proceed.}, for the magnetically dominated
force-free field, we arrive at
\begin{equation}\label{fffvirial}
M = \int_{\partial \mathrm{V}} \frac{\alpha^2 B^2}{2} \left( \mathbf{r\cdot} d \mathbf{S} \right) %
- \int_{\partial \mathrm{V}} \alpha^2 \left( \mathbf{B \cdot r}\right) %
\left(\mathbf{B\cdot}d \mathbf{S} \right) %
+ \int_{\mathrm{V}} \frac{(1 - \alpha^2)}{2} \left( B_r^2 + B^2
\right) dV \ .
\end{equation}
Assuming that the magnetic field vanishes sufficiently rapidly at
large distances, we find that the energy of the force-free fields
in the exterior $r>r_0$ of the neutron star is
\begin{equation}\label{Mdefinition2}
M = \pi r_0^3 \int \alpha^2 \left(B_r^2 - B_{\theta}^2\right)\bigg|_{r=r_0} \sin\theta d\theta %
+ \int_{\mathrm{V}} \frac{r_g}{r} \left(2 B_r^2 + B_{\theta}^2
\right) dV \ ,
\end{equation}
where $r_0$ is the radius of the neutron star.

We note that, in a flat spactime, the second term on the right
hand side of the above equation disappears and the total magnetic
energy of a force-free magnetic field in the exterior region
$r>r_0$ of a sphere is uniquely determined by the field values at
the the boundary $r=r_0$. However this is no longer the case for
the curved spacetime, since additional terms proportional to $r_g$
appear on the right hand side of this equation. Close observation
of equation (\ref{Mdefinition2}) shows that, when GR effects are
ignored, no force-free field that is completely detached from the
solar surface (i.e., $B_r = 0$ at $r = r_0$ in the exterior region
$r \ge r_0$) can exist (Low 2001). However, such completely
detached field configurations in the general relativistic magnetar
magnetosphere, due to the spacetime curvature, may be in the
equilibrium state\footnote{See the magnetic field configuration in
Figure 8b of Low (2001), which can not maintain equilibrium in
flat spacetime. But such configurations can be self-confined by
the spacetime curvature effects.}. This suggests that, besides the
normal flux at the magnetar surface, the GR spacetime curvature
provides additional self-confining effects. As a result, it needs
more work to be done to open the magnetic field in the curved
spacetime than in the flat spacetime. It is conceivable that when
the magnetar mass increases, this effect becomes more evident (see
Figure \ref{ratio}). Such GR effects have important implications
for the magnetic energy storage process in the magnetar
magnetosphere. In the next section, we will quantitatively
calculate their influences on the Aly-Sturrock constraint.





\section{Aly-Sturrock Constraint For Magnetic Field Energy}
To discuss the magnetic energy in the magnetar magnetosphere, it
is beneficial to introduce the potential field
$\mathbf{B}_{\mathrm{pot}}$ in the Schwarzschild metric which
satisfies (Uzdensky 2004)
\begin{equation}\label{fff}
\nabla \times (\alpha\mathbf{B}) = 0 \ ,
\end{equation}
and the boundary condition
\begin{equation}\label{bc1}
r = r_0 \ , B_r = F(\theta) \ .
\end{equation}
In this paper we mainly discuss the dipole field and its relevant
open state. The explicit expression of the dipole field can be
found in Appendix B. In this case the above boundary becomes $B_r
= C \cos\theta$, where $C$ is a constant. Note that the potential
field now involves the spactime curvature term $\alpha$ in
equation (\ref{fff}). This is quite different from the flat
spacetime definition (Komissarov 2004). Note that, as
$\alpha$$\rightarrow$$1$, this potential field definition reduces
to its flat spacetime form. The associated magnetic energy of the
potential field is designated as $M_{\mathrm{pot}}$. For the
force-free field in the magnetosphere, there exists one
interesting energy reference state, the Aly-Sturrock state (Aly
1984,1991; Sturrock 1991). Imagine all force-free magnetic fields
complying with the boundary condition (\ref{bc1}), with one end of
each line of force anchored to the star's surface and the other
out to infinity. Among all these fields, the one with the lowest
energy is potential everywhere except for a current sheet at the
equator (Aly 1984,1991; Sturrock 1991). This lowest energy state
is the Aly-Sturrock state.
Call this magnetic field configuration
$\mathbf{B}_\mathrm{open}$. 
The total energy of this state is designated as
$M_{\mathrm{open}}$. The well-known Aly-Sturrock conjecture claims
that for any fully closed force-free field\footnote{Strictly
speaking, this condition is not fulfilled since the field lines
open at the light cylinder. Fortunately, magnetars are slow
rotators, so the light cylinder is quite far away from the neutron
star surface and this effects can be negligible. For this reason,
we focus in this paper on the non-rotating neutron stars.} with
the boundary condition (\ref{bc1}),
its total energy $M_{\mathrm{FF}}$ satisfies the
following relation,
\begin{equation}\label{alysturrock}
M_{\mathrm{pot}} < M_{\mathrm{FF}} < M_{\mathrm{open}} \ .
\end{equation}
This first half of this inequality means a current-free potential
field is the lowest energy state. And the second half suggests
that the opening up process of an initial
closed force-free magnetic field 
requires considerable amount of work to be done on the magnetic
field.
Of particular interest is whether the pre-eruption magnetic energy
$M$ can exceed the threshold set by the Aly-Sturrock field. This
is crucial for the magnetically driven outbursts.

Some numerical experiments have recently demonstrated the validity
of this conjecture (Antiochos, DeVore \& Klimchuk 1999; Hu 2004).
Due to the importance of the Aly-Sturrock constraint for the
magnetic eruption, it is worthwhile to reconsider this problem
when GR effects are important. Note that this Aly-Sturrock state
is unique (Aly 1984; Sturrock 1991) and can be constructed by the
following technique. Modify the boundary condition (\ref{bc1}) to
\begin{equation}\label{bc2}
r = r_0 \ , B_r = |F(\theta)| \ .
\end{equation}
After getting the field with this boundary condition and reversing
the directions of those lines at the boundary $r=r_0$ where $B_r <
0$ of this field, we could get the Aly-Sturrock state (see also
Low \& Smith 1993).
We have calculated the fully open field
$\mathbf{B}_{\mathrm{open}}$ and the relevant energy
$M_{\mathrm{open}}$ numerically.
The details to obtain the Aly-Sturrock field and the magnetic
energy $M_{\mathrm{open}}$ are discussed in Appendix C. In Figure
\ref{Alyexample}, an illustrative example of the fully open
Aly-Sturrock field is shown. The current sheet at the equator is
shown by a thick solid line.

\begin{figure}
\begin{center}
\epsfig{file=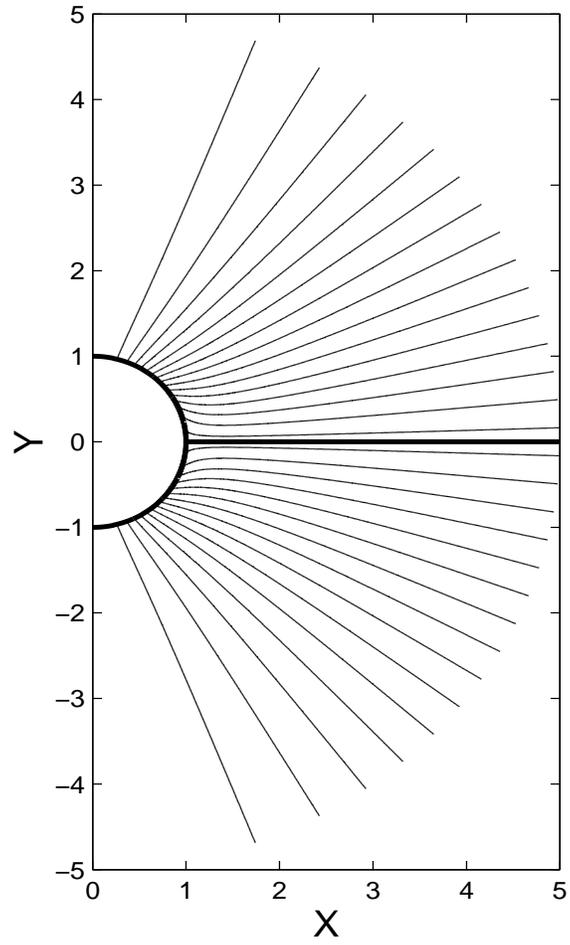,height=5in,width=3in,angle=0}
\end{center}
\caption{
         The fully open Aly-Sturrock field with
         an initial dipole field boundary condition.
         The thick solid line at the equator denotes
         the current sheet in the field.
         }
\label{Alyexample}
\end{figure}



\subsection{Dependence of $M_{\mathrm{open}}$
on Neutron Star Masses }

\begin{figure}
\begin{center}
\epsfig{file=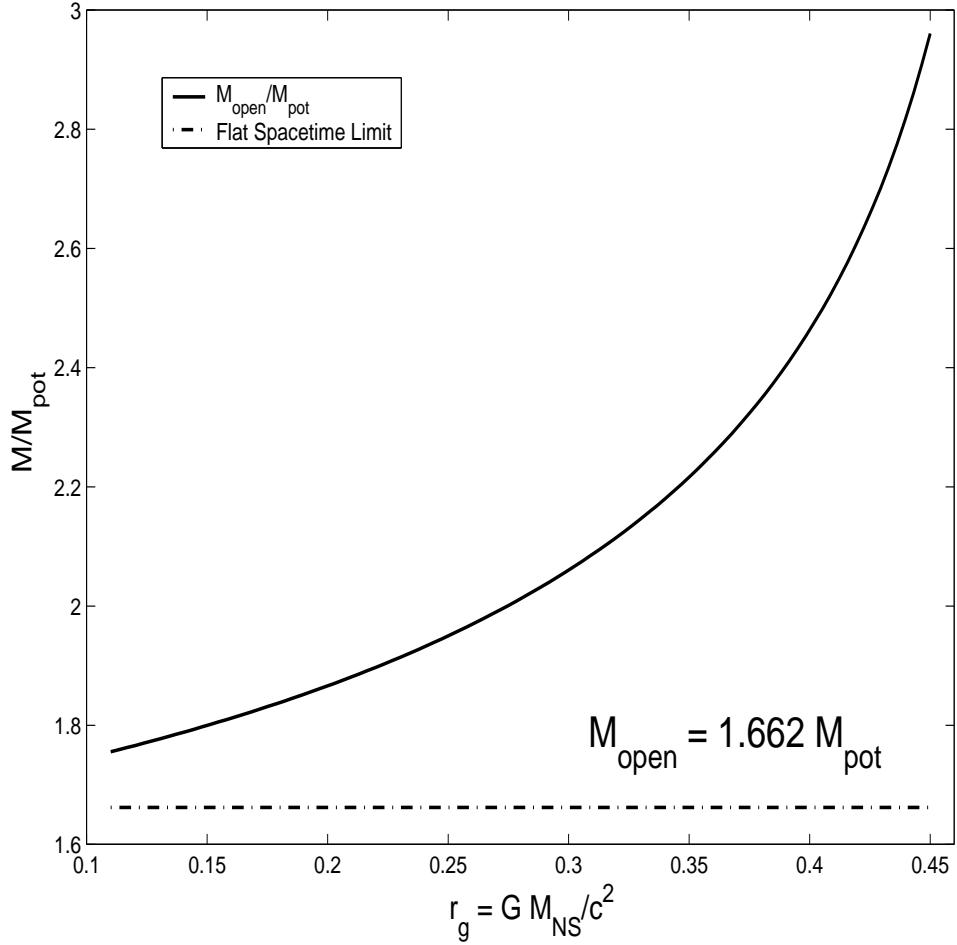,height=5in,width=5in,angle=0}
\end{center}
\caption{
         The variation of the $M_{\mathrm{open}}$
         (in units of $M_{\mathrm{pot}}$) with $r_g$
         (equivalently, the magnetar mass).
         Note $r_g $ ranges from 0.15 to 0.45, which corresponds
         to a mass range of $1-3M_{\odot}$.
         Note that in flat spacetime the fully open
         field energy $M_{\mathrm{open}} = 1.662 M_{\mathrm{pot}}$
         for the dipole field boundary condition,
         which is denoted by dot-dashed line in this figure.
         }
\label{ratio}
\end{figure}

To investigate the spacetime curvature effects on the Aly-Sturrock
constraint, we calculate the Aly-Sturrock threshold
$M_{\mathrm{open}}$ for different magnetar masses.
Throughout this paper we take the neutron star radius $r_0 = 1$,
so for a neutron star mass of $1-3$ $M_{\odot}$, $r_g$ ranges from
$0.15-0.45$ (For simplicity, we keep the neutron star radius fixed
at 10 $\mathrm{km}$, though this is not the case in reality). In
Figure \ref{ratio}, we show the variation of $M_{\mathrm{open}}$
(in units of $M_{\mathrm{pot}}$) with the neutron star mass. This
figure shows that the more massive the magnetar, the higher the
threshold is. For instance, for the dipole field with $r_g = 0.15$
(1 $M_{\odot}$), the energy of the fully open Aly-Sturrock field
is $M_{\mathrm{open}}$ = $1.80 M_{\mathrm{pot}}$; when $r_g =
0.21$ (1.4 $M_{\odot}$), the energy becomes $M_{\mathrm{open}}$ =
$1.88 M_{\mathrm{pot}}$. Consequently, it is more difficult for
more massive neutron stars to surpass the Aly-Sturrock energy
threshold. From this figure, we also note that as $r_g \rightarrow
0$ the Aly-Sturrock threshold approaches the flat spacetime limit
$M_{\mathrm{open}} = 1.662 M_{\mathrm{pot}}$. This is consistent
with our physical expectation.

This increase with mass of the Aly-Sturrock energy threshold stems
entirely from the spacetime curvature self-confining effects
mentioned in Section 2 and this behavior is quite different from
the solar eruption in flat spacetime, in which the Aly-Sturrock
field energy ($\sim$ 1.662 $M_{\mathrm{pot}}$) is independent of
the star mass.
It should be emphasized that in the magnetar outbursts, the
Aly-Sturrock energy constraint is more stringent than for the
solar CME-type eruptions. From Figure 2, we can infer that
magnetars are probably not neutron stars with extreme mass $\sim 3
M_{\odot}$, as the Aly-Sturrock threshold could hardly be reached.

For typical neutron star masses ($\sim 1-2 M_{\odot}$, $r_g \sim
0.15 - 0.3$),
it is necessary to seek initial magnetic fields which possess
magnetic energy in excess of the threshold set by the Aly-Sturrock
energy $M_{\mathrm{open}}$.
One possibility is due to the mass loading effects. The estimated
ejected mass loading is about $10^{22}\mathrm{g}$ (Lyutikov 2006).
This mass loading can be balanced by pressure forces in the
vertical direction. The pressure gradient in the horizontal
direction, however, requires magnetic forces associated with
cross-field currents, i.e., $\mathbf{J} \times \mathbf{B} \neq 0$,
to maintain the equilibrium state. The deviations from a strictly
force-free magnetic fields, i.e., the cross-field contribution,
are worth further investigations (Low \& Smith 1993, Wolfson \&
Dlamini 1997).
Physically speaking, the mass loading would act as a lid over the
magnetic field.
The field can be compressed globally by a sufficient amount of
plasma. As a result, the energy of the compressed magnetic field
increases as the total load increases, and eventually the magnetic
energy exceeds $M_\mathrm{open}$. In other words, the cross field
current densities provide additional sources of magnetic free
energy which may be enough to enable the magnetic field to clear
the
threshold $M_\mathrm{open}$.  %






\section{Axisymmetric Magnetostatic Magnetosphere with Cross-Field Currents}
In this section we explore the cross-field effects on the magnetic
energy storage properties in the magnetar magnetosphere. Similar
investigations in solar CMEs have been carried out by Zhang \& Low
(2004). Specifically, we will focus on the question whether the
magnetic energy in the magnetosphere can exceed the Aly-Sturrock
threshold. In what follows, we consider that the magnetar
magnetosphere evolves quasi-statically on sufficiently slow
timescale that we can treat the magnetosphere as being essentially
in magnetostatic equilibrium.
A steady state axisymmetric, purely poloidal magnetic field in the
Schwarzschild metric can be written as
\begin{equation}
{\bf B} = {\bf B}_{\rm pol} = \nabla \Psi \times \nabla\phi \ ,
\end{equation}
where $\Psi(r,\theta)$ is the poloidal magnetic stream function.
The ``ordinary" orthogonal basis is used, where ${\bf
e}_{\hat{\mu}} = g_{\mu\mu}^{-1/2}\partial_{\mu}$ (Weinberg 1972,
no summation rule over $\mu$ is used in this equation), namely,
\begin{equation}
{\bf e}_{\hat{r}} = \alpha\partial_r \ , {\bf e}_{\hat{\theta}} =
\frac{1}{r}\partial_{\theta} \ , {\bf e}_{\hat{\phi}} =
\frac{1}{r\sin\theta}\partial_{\phi} \ .
\end{equation}
The poloidal magnetic field components are (Uzdensky 2004)
\begin{equation}\label{brbt}
{\bf B} = \frac{1}{r\sin\theta} \left( \frac{1}{r}\frac{\partial
\Psi}{\partial \theta} , \  - \alpha \frac{\partial \Psi}{\partial
r} \right) \ . 
\end{equation}
To account for the the cross-field current effects induced by the
mass loading, we must go beyond the force-free approximations (Yu
2011) and turn to the full magnetohydrodynamic (MHD) equation
(\ref{derivevirial}).
This equation decomposes into the following two equations
\begin{equation}\label{gs}
\frac{\partial }{\partial r}\left(\alpha^2 \frac{\partial \Psi}{\partial r} \right) %
+ \frac{\sin\theta}{r^2}\frac{\partial}{\partial \theta}%
\left( \frac{1}{\sin\theta}\frac{\partial \Psi}{\partial\theta} \right) %
+ r^2\sin^2\theta\frac{\partial p}{\partial \Psi} = 0 \ ,
\end{equation}
\begin{equation}\label{equili2}
g^{rr}\frac{\partial p}{\partial r} %
+ ( p + \rho) \frac{G \mathcal{M}_{\mathrm{ns}}}{r^2} = 0 \ , %
\end{equation}
for balance across and along the magnetic field (Low \& Smith
1993). 
A simple solution to equation
(\ref{equili2}) reads
\begin{equation}\label{linear1}
p = \frac{P(\Psi)}{r^{m+1}} \ , 
\end{equation}
\begin{equation}\label{linear2}
\rho_0 = \frac{1}{G\mathcal{M}_{\mathrm{ns}}} \frac{P(\Psi)}{r^{m}}\left[ m+1 - %
\left(2m + 2 + \frac{\gamma}{\gamma-1} \right)\frac{r_g}{r} \right] \ , %
\end{equation}
where $P(\Psi)$ is a free function of the magnetic stream function
$\Psi$ and $m$ is a constant.

To keep the problem mathematically tractable, we take the free
function $P$ to be linear in $\Psi$. Subsequently, equation
(\ref{linear1}) and (\ref{linear2}) become
\begin{equation}\label{linear3}
p = \frac{\lambda (\Psi + \Psi_0) }{r^{m+1}} \ , %
\end{equation}
\begin{equation}\label{linear4}
\rho_0 = \frac{1}{G\mathcal{M}_{\mathrm{ns}}} \frac{\lambda(\Psi + \Psi_0)}{r^{m}}\left[ m+1 - %
\left(2m + 2 + \frac{\gamma}{\gamma-1} \right)\frac{r_g}{r} \right] \ , %
\end{equation}
where $\Psi_0$ and $\lambda$ are constants. Substitute equation
(\ref{linear3}) into equation (\ref{gs}), we obtain the following
linear Grad-Shafranov equation
\begin{equation}
\frac{\partial }{\partial r}\left(\alpha^2 \frac{\partial \Psi}{\partial r} \right) %
+ \frac{\sin\theta}{r^2}\frac{\partial}{\partial \theta}%
\left( \frac{1}{\sin\theta}\frac{\partial \Psi}{\partial\theta} \right) %
+ \lambda \frac{\sin^2\theta}{r^{m-1}} = 0 \ .
\end{equation}
The general solution to the above equation can be written as
\begin{equation}\label{casestudy}
\Psi = f_m(r)\sin^2\theta + \Psi_{\mathrm{pot}} \ ,
\end{equation}
where $\Psi_{\mathrm{pot}}$ is an arbitrary potential stream
function satisfying (Ghosh 2000)
\begin{equation}
\frac{\partial }{\partial r}\left(\alpha^2 \frac{\partial \Psi_{\mathrm{pot}}}{\partial r} \right) %
+ \frac{\sin\theta}{r^2}\frac{\partial}{\partial \theta}%
\left( \frac{1}{\sin\theta}\frac{\partial \Psi_{\mathrm{pot}}}{\partial\theta} \right) %
 = 0 \ .
\end{equation}
This equation can be readily solved by the variable separation
method (see Appendix B). The function $f_m(r)$ satisfies the
following second order ordinary differential equation (ODE)
\begin{equation}\label{ODE}
\left( 1 - \frac{2 r_g}{r} \right) f^{\prime\prime} + \frac{2
r_g}{r^2} f^{\prime} - \frac{2}{r^2} f + \frac{\lambda}{r^{m-1}} =
0 \ ,
\end{equation}
where prime denotes derivatives with respect to $r$.
The particular solutions can be readily obtained analytically.
For $m=3, 4, 5, 6, 7,$ and $8$, the radial function $f_m$ are
given explicitly in Appendix D. The simple linear solutions given
by the above equations can not be expected to describe the
magnetar magnetosphere in realistic details. However, the solution
given by equation (\ref{casestudy}) with $\Psi_{\mathrm{pot}} = 0$
can be used to obtain a physical estimate of how much energy can
be stored in the magnetosphere prior to eruptions.


The magnetic energies for different values of $m$ and $r_g$ are
listed in Table 1. We find that, for $r_g = 0.15$ and $0.21$ , the
field configurations are able to sustain magnetic energy higher
than the Aly-Sturrock threshold as $m \ge 8$.
\begin{table} 
  \caption{Values of $M/M_{\mathrm{pot}}$ for different values of
  $m$.   }
  \label{table-1}
  \begin{center}\begin{tabular}{cccccccc}
  \hline\noalign{\smallskip}
$m$ &  $r_g = 0.15$          & $r_g = 0.21 $       & $r_g = 0.3      $ \\
  \hline\noalign{\smallskip}
3  & 2.19 & 2.30 & 2.55 \\
4  & 1.00 & 1.00 & 1.00 \\
5  & 1.15 & 1.13 & 1.10 \\
6  & 1.44 & 1.38 & 1.30 \\
7  & 1.77 & 1.67 & 1.52 \\
8  & 2.12 & 1.98 & 1.76 \\
9  & 2.48 & 2.29 & 2.00 \\
10 & 2.84 & 2.61 & 2.25 \\
  \noalign{\smallskip}\hline
  \end{tabular}\end{center}
The gravitational radius $r_g$ is taken as $0.15$, $0.21$ and
$0.3$, which correspond to magnetar mass of $1.0$, $1.4$ and $2.0$
$M_{\odot}$. The Aly-Sturrock energy thresholds, shown in Figure 2
for the three values of the magnetar mass, are $M_{\mathrm{open}}
= 1.80 M_{\mathrm{pot}}$, $M_{\mathrm{open}} = 1.88
M_{\mathrm{pot}}$ and $M_{\mathrm{open}} = 2.06 M_{\mathrm{pot}}$,
respectively. According to this table, we note that, when $m \ge
8$ for $r_g = 0.15, 0.21$ and $m \ge 10$ for $r_g = 0.3$, the
magnetic energy in the magnetosphere could be higher than the
Aly-Sturrock threshold.
\end{table}
Simple estimation shows that the total magnetic energy of a
magnetar with magnetic field $\sim 10^{14} - 10^{15}$ G is
approximately $ 10^{46}-10^{48}$ ergs. Given the actual giant
flare energy release $\sim 10^{44} - 10^{46}$ ergs, we know that a
few percent of the magnetic energy in excess of the Aly-Sturrock
threshold is needed to release during a giant flare. This energy
requirement can be fulfilled as $m$ reaches a critical value. For
instance, we find that for $r_g = 0.15$ the magnetic energy $M$
with $m = 8$
is approximately 15 percent above the Aly-Sturrock threshold
$M_{\mathrm{open}}$, which is enough to drive magnetar giant
flares. In Figure 3, we show the $m=8$ solution with $r_g = 0.15$
\[
\Psi = \frac{f_8(r)}{f_8(r_0)}\sin^2\theta \ , \ %
\rho_0 = \frac{1}{G\mathcal{M}_{\mathrm{ns}}} %
\frac{\lambda \Psi }{r^{m}}\left[ m+1 - %
\left(2m + 2 + \frac{\gamma}{\gamma-1} \right)\frac{r_g}{r}
\right] \ ,
\]
where the stream flux is normalized to unity at $r=r_0$ and
$\theta = \pi/2$ and we have set the constant $\Psi_0 = 0 $ in
equation (\ref{linear4}). The left panel in this figure shows the
magnetic field lines and the right one shows the contour of the
density departure\footnote{Note that an arbitrary, spherically
symmetric density distribution corresponds to the term that is
proportional to the constant $\Psi_0$ in equation (\ref{linear4}),
which is ignored in this figure.} from an arbitrary, spherically
symmetric distribution. In this particular state, the magnetic
energy $M$ is $2.12 M_{\mathrm{pot}}$, which is greater than the
Aly-Sturock state $M_{\mathrm{open}} = 1.80 M_{\mathrm{pot}}$.

The solution with $m=3$ and $\Psi_{\mathrm{pot}} = 0$ is a purely
radial magnetic field,
\[ 
B_r = \lambda \frac{\cos\theta}{r^2} \ , B_{\theta} = 0 \ .
\] 
This solution has been extensively discussed in the
Blandford-Znajek process (e.g. Blandford \& Znajek 1977) related
to relativistic astrophysical jets. But in our discussion this
state itself is of no particular interest as we are more concerned
with the initial closed state. To introduce the closed field
structures,
\begin{figure}
\begin{center}
\epsfig{file=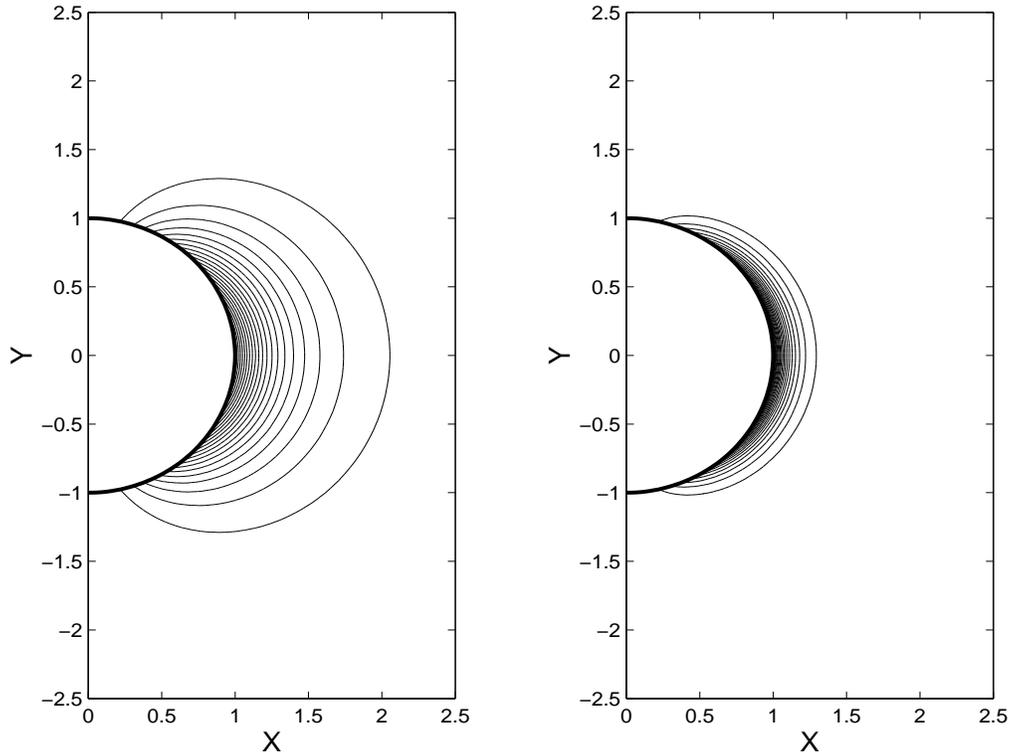,height=4in,width=5.3in,angle=0}
\end{center}
\caption{
         The $m=8$ solution with $r_g = 0.15$. The left panel
         shows the magnetic field. The right panel
         shows the density departure
         from an arbitrary, spherically
         symmetric distribution, i.e., we set
         the constant $\Psi_{0} = 0 $ in equation
         (\ref{linear4}).
         The energy of this state is $2.12 M_{\mathrm{pot}}$,
         which exceeds the Aly-Sturrock threshold $1.80 M_{\mathrm{pot}}$.
         }
\label{meq7}
\end{figure}
we add a dipole field to the $m=3$ purely radial magnetic field,
i.e.,
\begin{equation}\label{m3mix}
\Psi = \lambda \frac{f_3(r)}{f_3(r_0)} \sin^2\theta \pm
\Psi_{\mathrm{dipole}} \ ,
\end{equation}
where the stream function is also normalized. The magnetic fields
with $r_g = 0.15$ are shown in Figure 4. The left panel in this
figure corresponds to the ``$+$" sign, which approximately models
the effects of the neutron star wind (Bucciantini et al. 2006).
Such configurations are also discussed by Low \& Tsinganos (1986)
and applied to model the effects of solar wind. Note that when
$\lambda$ increases to $5.0$, the magnetic energy in the left
panel is about $1.83 M_{\mathrm{pot}}$, exceeding the
corresponding Aly-Sturock energy $M_{\mathrm{open}}$ by about
$2\%$, which suggests this state may support a giant flare. If
$\lambda$ is even increased, more magnetic energy can be obtained.
The right panel takes the ``$-$" sign in the above equation.
Though the right panel shows a state that is physically
unacceptable, it is worth pointing out that the energy of the
state with detached field lines ($\sim 2.86 M_{\mathrm{pot}}$) is
much higher than the energy in the left panel. This also suggests
that, when there are detached fields in the magnetosphere, the
stored magnetic energy can be much larger than those
configurations whose field lines are all anchored to the magnetar
surface.  This possibility to bypass the Aly-Sturrock constraint
has been discussed by Flyer et al. (2004) for solar CMEs and will
be further discussed for magnetar giant flares (Yu et al. in
prep).
\begin{figure}
\begin{center}
\epsfig{file=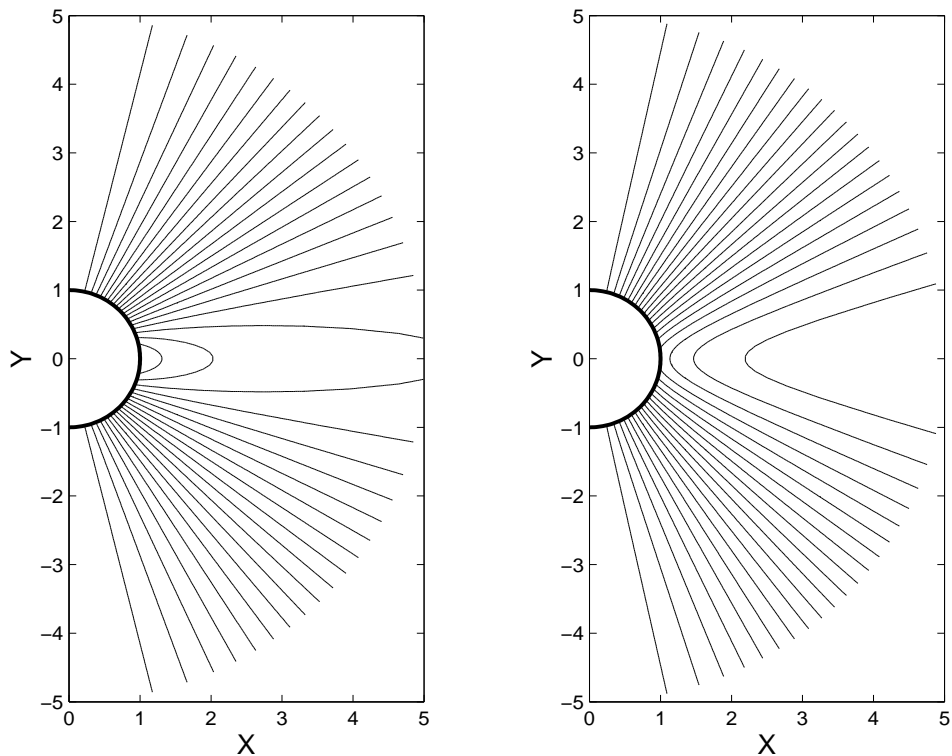,height=4in,width=5in,angle=0}
\end{center}
\caption{
         The magnetic fields in equation (\ref{m3mix})
         with $\lambda = 5.0$ and $r_g = 0.15$. The left panel corresponds
         to the ``$+$" sign and the right the ``$-$" sign.
         The energies of the two state are $1.83$ and $2.86$
         $M_{\mathrm{pot}}$, respecitvely. Both of them exceed the
         Aly-Sturrock threshold $M_{\mathrm{open}} = 1.80 M_{\mathrm{pot}}$.
         Though the state in the right panel  is physically
         unacceptable, it is instructive that
         configurations with detached field lines
         can build up more energy than the simple
         connected field lines.
         }
\label{meq3potimposed}
\end{figure}





\section{Conclusions and Discussions}\label{sec:diss}
We construct general relativistic models of non-rotating neutron
stars endowed with strong magnetic fields. The equilibrium states
of axisymmetric force-free magnetic fields in magnetar
magnetospheres are found as solutions of the Grad-Shafranov
equations in the Schwarzschild geometry. A newly derived general
relativistic magnetic virial theorem in presented in this work.
Based on this magnetic virial theorem, we carefully examine the GR
effects on the well known Aly-Sturrock energy threshold. We found
that this energy threshold increases with the magnetar mass. As a
result, it is more difficult for massive magnetars to erupt. By
this observation, we conclude that magnetars are probably not
neutron stars with extreme mass. The non-force-free magnetic field
induced by the mass loading is further investigated as a
possibility to bypass the Aly-Sturrock constraint for typical
magnetar mass around $\sim 1.4 M_{\odot}$.



We mainly discuss dipolar surface boundary conditions in this
paper. This is the case for magnetar's large scale fields. But
observations show a striking feature that the emergence of a
strong four-peaked pattern in the light curve of the 1998 August
27 event from SGR 1900+14, which was shown in data from the
Ulysses and Beppo-SAX gamma-ray detectors (Feroci et al. 2001).
These remarkable data may imply that the geometry of the magnetic
field was quite complicated in regions close to the star where GR
effects are important.
As a result, complex boundary conditions should be important for
the outburst of magnetars. Effects of different boundary
conditions on the energy buildup in magnetars are worth further
investigations (Antiochos et al. 1999).

For simplicity, we have neglected the relativistic wind from the
neutron star surface. Actually, the wind from the neutron star
(e.g. Bucciantini et al. 2006) may cause part of the magnetic
field lines to be in the open states before eruption. Similar
effects have been explored in solar CMEs (Low \& Smith 1993;
Wolfson 1993). It is interesting to investigate the effects of
neutron star wind on the magnetic energy storage properties.

Helicity has been discussed extensively in solar physics (Zhang \&
Low 2005). CMEs are believed to be the unavoidable products of the
coronal evolution as a result of magnetic helicity accumulation
(Zhang et al. 2006). But helicity in the GR regime is not a
well-explored issue. Finding a self-consistent definition of
helicity in the curved spactime and investigating the relevant
helicity properties are interesting topics for further
explorations.


The field topology change from a closed state to an open state
must be accompanied by the magnetic reconnection. After a certain
threshold is reached, the dynamical instability sets in. The
gradual quasi-static evolution of the magnetar's magnetosphere
will be replaced by the dynamical evolution of the field. This
naturally explains the problem as to how a very slow buildup of
the external shear (over an interval of $\sim$100 yr) could lead
to the sudden release of external magnetic energy on a much
shorter timescale (Lyutikov 2006). The magnetic energy dissipation
in the strongly magnetized plasma is caused by the tearing mode
instability (Lyutikov 2003, Komissarov et al. 2007). Relativistic
tearing instability induced reconnections in the nonlinear regime
need further studies to better understand the magnetar outburst
behaviors.

Our theoretical models can not address the nonlinear dissipation
processes that occur during giant flares. However, current GRMHD
simulations provide a unique opportunity to study the dynamical
outburst physics. The models constructed in this work are likely
to be useful as initial states in GRMHD numerical simulations to
explore the dynamics of magnetic eruptions (Gammie et al. 2003, Yu
2011).









\acknowledgments

We thank the anonymous referee for important comments and
suggestions that improve this paper greatly. The research is
supported by the Natural Science Foundation of China (Grant
10873033, 10703012, 10778702 and 10973034), the Western Light
Young Scholar Program and the 973 Program (Grant 2009CB824800).
The computation is performed at HPC Center, Kunming Institute of
Botany, CAS, China.

\clearpage


\appendix
\section{Derivation of Virial Theorem in Schwarzschild Metric}
The four equations expressing conservation of energy momentum are
\begin{equation}
\nabla_{\nu} T^{\mu\nu} = 0 \ , 
\end{equation}
where the Einstein summation rule is assumed and Greek letters
take on the values $t$, $r$, $\theta$, and $\phi$. The four
velocity for a plasma in magnetostaic equilibrium is
\begin{equation}
u^{t} = (- g_{tt})^{-1/2}, \  u^r=u^{\theta}=u^{\phi} = 0 \ .
\end{equation}
Given the energy-momentum tensor in equation
(\ref{energymomentum}), the covariant derivative can be expanded
as follows,
\[
\nabla_{\nu} T^{\mu\nu} = g^{\mu\nu}\frac{\partial}{\partial x^{\nu}}\left(p+\frac{b^2}{2}\right) %
+ \Gamma^{\mu}_{\beta\sigma}(p + \rho + b^2)u^{\beta} u^{\sigma}  %
\]
\begin{equation}
- \frac{1}{\sqrt{-g}}\frac{\partial (\sqrt{-g}\ b^{\mu} b^{\nu})}{\partial x^{\nu}}  %
- \Gamma^{\mu}_{\beta\sigma} b^{\beta} b^{\sigma} \ .
\end{equation}
The radial component of the above equation becomes (note that the
connection coefficients $\Gamma^{\mu}_{tt} = -\frac{1}{2}
g^{\mu\nu}\frac{\partial g_{tt}}{\partial x^{\nu}} $)
\[
g^{rr} \frac{\partial }{\partial r}\left( p + \frac{b^2}{2}\right) %
+ g^{rr}\frac{1}{2 g_{tt}}\left( p + \rho + b^2 \right) \frac{\partial g_{tt}}{\partial r} %
- \left( \frac{1}{\sqrt{-g}} \frac{\partial }{\partial r}\left(\sqrt{-g}\ b^r b^r \right) %
\right.
\]
\begin{equation} \label{radial}
\left.
+ \frac{1}{\sqrt{-g}} \frac{\partial }{\partial \theta}\left(\sqrt{-g}\ b^r b^{\theta} \right) %
+ \Gamma^r_{rr} b^r b^r %
+ \Gamma^r_{\theta\theta} b^{\theta} b^{\theta} %
+ \Gamma^r_{\phi\phi} b^{\phi} b^{\phi} \right) = 0 \ ,
\end{equation}
where
\[ \Gamma^r_{rr} = - \frac{r_g}{r(r - 2 r_g )} \ , \ %
\Gamma^r_{\theta\theta} = - (r - 2 r_g) \ , \ \Gamma^r_{\phi\phi} = %
- (r - 2 r_g) \sin^2\theta \ .
\]
The ``ordinary" component of the magnetic field $\mathbf{B}$ in
the orthogonal basis (Weinberg 1972) is related to the magnetic
field 4-vector $b^{\mu}$ by
\begin{equation}
B_r = \sqrt{g_{rr}} \ b^{r} = \sqrt{g^{rr}} \ b_{r} \ , %
B_{\theta} = \sqrt{g_{\theta\theta}} \ b^{\theta} = \sqrt{g^{\theta\theta}} \ b_{\theta} \ , %
B_{\phi} = \sqrt{g_{\phi\phi}} \ b^{\phi} = \sqrt{g^{\phi\phi}} \ b_{\phi} \ , %
\end{equation}
and
\begin{equation}
b^2 = b_{\mu} b^{\mu} = B^2=B_r^2 + B_{\theta}^2 + B_{\phi}^2 \ .
\end{equation}
Multiplying the equation (\ref{radial}) by $r$ and expressing the
magnetic field 4-vector $b^{\mu}$ by the ``ordinary" magnetic
fields $\mathbf{B}$ in equation (\ref{radial}), we may arrive at
\[
\alpha^2 r \frac{\partial }{\partial r}\left( p + \frac{B^2}{2}\right) %
+ \frac{r_g}{ r}\left( p + \rho + B^2 \right) %
- \frac{1}{r^2}\frac{\partial}{\partial r}\left( r^3 \alpha^2 B_r^2 \right) %
- \frac{\alpha}{\sin\theta}\frac{\partial}{\partial \theta}\left( \sin\theta B_r B_{\theta} \right) %
\]
\begin{equation}
+ \frac{r_g}{ r} B_r^2 + \alpha^2 (B_r^2 + B_{\theta}^2+
B_{\phi}^2) = 0 \ .
\end{equation}
Performing the volume integral with the usage of Gauss's theorem,
the above equation can be re-arranged to give the generalized
virial theorem, equation (\ref{virial}) in the main text.


\section{ Dipole Field Boundary Conditions and
Separable Solutions for Potential Fields}
To get the dipole field boundary conditions, we need to obtain the
current-free potential field. To be self-contained, we describe
the separable solutions of the homogenous Grad-Shfranov equation,
which are also the building blocks for the Aly-Sturrock fully open
field. The homogenous GS equation reads
\begin{equation}
\frac{\partial}{\partial r} \left[ \left( 1 - \frac{2 r_g}{r}
\right) \frac{\partial \Psi}{\partial r} \right] +
\frac{\sin\theta}{r^2}\frac{\partial }{\partial\theta}
\left(\frac{1}{\sin\theta} \frac{\partial
\Psi}{\partial\theta}\right) = 0 \ . \label{potential}
\end{equation}
Separable solutions of the above equation are of the form
\begin{equation}
\Psi(r,\theta) = R(r)\Theta(\theta) \ .
\end{equation}
Substitute the above equation into equation (\ref{potential}), we
obtain
\begin{equation}\label{theta}
\frac{d}{d\theta}\left( \frac{1}{\sin\theta} \frac{d \Theta}{d
\theta}\right) = - \lambda \frac{\Theta}{\sin\theta} \ ,
\end{equation}
\begin{equation}\label{R}
\frac{d}{d r}\left[ \left( 1- \frac{2 r_g}{r} \right) \frac{d R}{d
r}\right] = \lambda \frac{R}{r^2} \ ,
\end{equation}
where $\lambda$ is the separation constant. The lowest order of
solution are the special case with $\lambda = 0$, which can be
obtained by setting $\lambda = 0$ in the above two equations. The
solutions are then
\begin{equation}
\Theta(\theta) = {\rm a} \cos\theta + {\rm b} \ ,
\end{equation}
\begin{equation}
R(r) = {\rm c} \ ,
\end{equation}
where a,b, and c are constants. This solution 
is the Schwarzschild monopole.

The order of the solution is denoted by the ordinal number $m$
($m=1$ corresponds to the dipole field), related to the constant
$\lambda$ by $\lambda = m(m+1)$. Equations (\ref{theta}) and
(\ref{R}) become (Ghosh 2000):
\begin{equation}
(1 - \mu^2) \frac{d^2 \Theta}{d\mu^2} + m(m+1) \Theta = 0 \ ,
\label{mu}
\end{equation}
\begin{equation}
(1 - z^2) \frac{d^2 R}{dz^2} - 2 \frac{d R}{d z}+ m(m+1) R = 0 \ ,
\label{Jacobi}
\end{equation}
where $\mu = \cos\theta$ and $z = r/r_g - 1$.

The solution of equation (\ref{mu}) is
\begin{equation}
\Theta (\mu) = (1 - \mu^2) \frac{d P_{m}(\mu)}{d \mu} \ ,
\end{equation}
where $P_{m}(\mu)$ is the Legendre polynomial. The solutions of
equation (\ref{Jacobi}) are
\begin{equation}
R(r) = r^2 \left\{ \begin{array}{l}
{\cal P}^{(0,2)}_{m-1}(z) \\
{\cal Q}^{(0,2)}_{m-1}(z) \\
\end{array} \right.  \ ,
\end{equation}
where ${\cal P}^{(0,2)}_{m-1}(z)$ and ${\cal Q}_{m-1}^{(0,2)}(z)$
are Jacobi polynomial  and Jacobi functions of the second kind,
respectively. For $r\gg r_g$, the Jacobi polynomial and Jacobi
function's asymptotic behaviors are (Szeg$\mathrm{\ddot{o}}$ 1939)
\begin{equation}
{\cal P}^{(0,2)}_{m-1}(z) \sim r^{m-1}  \ , \ {\cal
Q}^{(0,2)}_{m-1}(z) \sim r^{-m-2} \ .
\end{equation}
The superscripts in the Jacobi polynomial and Jacobi function will
be suppressed hereafter, as the values remain the same throughout
this study. The explicit expressions for the Jacobi polynomials
and Jacobi functions can be found in Gradshteyn \& Ryzhik (1980).

Of particular interest is the dipole configuration ($m=1$)
determined by
\begin{equation}
\Psi =  \left[(1-\mu^2)\frac{P_1(\mu)}{d\mu}\right] r^2 {\cal
Q}_0(z) =  \left[ \frac{r^2}{2}\ln\left(\frac{r}{r - 2 r_g}\right)
- r r_g - r_g^2 \right] \sin^2\theta \ .
\end{equation}
This solution can be used as boundary conditions. 

\section{Determination of the Aly-Sturrock Field}
To appreciate the Aly-Sturrock constraint on the availability of
free magnetic energy, we need to determine the Aly-Sturrock state
numerically. Following Low \& Smith (1993),
%
%
the boundary conditions of Aly-Sturrock fully opened field can be
obtained by flipping the flux function according to the boundary
condition (\ref{bc2})
\begin{equation}
\Psi_{\mathrm{modify}} = \left\{ \begin{array}{ll}
\Psi(r_0, \theta)                    & 0\leq \theta \leq \pi/2 \\
2\Psi(r_0,\pi/2) - \Psi(r_0, \theta) & \pi/2\leq \theta \leq \pi \\
\end{array} \right.  \ .
\end{equation}
Specifically, for the original dipole boundary condition, the
modified boundary condition becomes
\begin{equation}\label{mbc}
  \Psi_{\mathrm{modify}}(r_0,\theta) = B_0 A_1 \times \left\{\begin{array}{ll}
       \sin^2\theta                      &   0\leq \theta \leq \pi/2 \\
       2 - \sin^2\theta                  &   \pi/2\leq \theta \leq \pi \\
       \end{array} \right.  \ ,
\end{equation}
where
\[
A_1 = \frac{r_0^2}{2}\ln\left(\frac{r_0}{r_0 - 2 r_g} \right) -
r_0 r_g - r_g^2 \ ,
\]
and $r_0$ is the magnetar radius. The solutions to the homogeneous
Grad-Shafranov equation are of the form
\begin{equation}
\Psi(r, \theta) = \sum_{n=1}^{\infty} a_{n} \left(r^2 {\cal
Q}_{n-1}(r) \right) \left( \sin^2\theta \frac{d P_{n}(\mu)}{d\mu}
\right) + \alpha_0 + \alpha_1 \cos\theta  \ ,
\end{equation}
where $\mu = \cos\theta$ and $\mathcal{Q}_{n-1}(r)$ is the Jacobi
function of the second kind. It is clear that
\begin{equation}
\alpha_0 = B_0 A_1 \ , \  \alpha_1 = - B_0 A_1 \ .
\end{equation}
We define the following flux function as
\[
  \Psi^{*}(r, \theta) = \Psi(r,\theta) - \alpha_0 - \alpha_1
  \cos\theta \ .
  \]
The problem becomes to determine the coefficient $a_n$
\begin{equation}\label{psidcmp}
\Psi^{*}(r, \theta) = \sum_{n=1}^{\infty} a_{n} \left(r^2 {\cal
Q}_{n-1} \right) \left( \sin^2\theta \frac{d P_{n}(\mu)}{d\mu}
\right) \ ,
\end{equation}
subject to the modified boundary condition (\ref{mbc})
\[
  \Psi^{*}(r_0, \theta) = \Psi(r_0,\theta) - \alpha_0 - \alpha_1 \cos\theta
  \]
\begin{equation}
=  B_0 A_1 \times \left\{\begin{array}{ll}
       \sin^2\theta -1 + \cos\theta                     &   0\leq \theta \leq \pi/2 \\
       1 - \sin^2\theta + \cos\theta                 & \pi/2\leq \theta \leq \pi \\
       \end{array} \right.  .
\end{equation}
According to the orthogonality of associated Legendre polynomials
$ P_{n}^{1}(\mu) $, we have that
\begin{equation}
a_n = - \frac{1}{r_0^2 {\cal Q}_{n-1}(r_0)} \frac{2n+1}{2n(n+1)}
\int^{\pi}_0 \Psi^{*}(r_0,\theta) P_{n}^{1}(\mu) d\theta \ ,
\end{equation}
Note that $\Psi^{*}(r_0,\theta)$ is an odd function of $\theta$ in
the integration range. When $n$ is an odd integer, the
coefficients $a_n$'s vanish. The non-zero coefficients $a_{n}$'s
($n=2m$) can be written as
\begin{equation}
a_{2m} = - \frac{B_0 A_1} {r_0^2 {\cal Q}_{2m-1}(r_0)  }
\frac{4m+1}{2m(2m+1)} \int^{\pi/2}_{0} (\sin^2\theta -1 +
\cos\theta) P_{2m}^1(\cos\theta) d\theta  \ , 
\end{equation}
where $P_{2m}^1(\cos\theta)$ is the associated Legendre
polynomial. After some manipulations, we arrive at
\begin{equation}
a_{2m} = \frac{B_0 A_1} {r_0^2 {\cal Q}_{2m-1}(r_0)  }
\frac{4m+1}{m(2m+1)} \frac{(-1)^{m-1}(2m-2)!}{2^{2m}(m-1)!(m+1)!}
\equiv \frac{c_{2m}}{{\cal Q}_{2m -1}(r_0) }\ .
\end{equation}
The radial and $\theta$ components of the magnetic field,
according to equation (\ref{brbt}), are
\begin{equation}
B_r = \sum_{m=1}^{\infty} a_{2m} {\cal Q }_{2m-1}(r) \left[ 2m
(2m+1) P_{2m}\right] - \frac{\alpha_1}{r^2} \ ,
\end{equation}
and
\begin{equation}
B_{\theta} = -\sqrt{1 - \frac{2 r_g}{r}} \sum_{m=1}^{\infty}
a_{2m} \left[ 2{\cal Q }_{2m-1}(r) + r {\cal Q
}^{\prime}_{2m-1}(r) \right] \sin\theta \frac{ d P_{2m}(\mu)
}{d\mu} \ ,
\end{equation}
where prime denotes derivative with respect to $r$. The magnetic
energy of the open field, according to equation
(\ref{Mdefinition2}), is
\[
M_{\mathrm{open}} = \pi r_0^3 \left(1 - \frac{2 r_g}{r_0} \right)%
\left\{2 B_0^2 \frac{A_1^2}{r_0^4} + \sum_{m=1}^{\infty} c_{2m}^2
\frac{4m(2m+1)}{4m+1} \times \right.
\]
\[
\left. \left[ 2m(2m+1) - \left( 1 - \frac{2 r_g}{r_0} \right)
\left( 2 + \frac{ r_0 {\cal Q }^{\prime}_{2m-1}(r_0)}{{\cal
Q}_{2m-1}(r_0)}\right)^2 \right] \right\} +
\]
\[
2 \left( 4\pi r_g B_0^2 A_1^2 \int_{r_0}^{\infty} %
\frac{1}{r^3} dr %
+ 4\pi r_g \sum_{m=1}^{\infty} c_{2m}^2 \frac{\left[2m(2m+1)\right]^2}{4m+1}  \int_{r_0}^{\infty}  %
\left(\frac{\mathcal{Q}_{2m-1}(r)}{\mathcal{Q}_{2m-1}(r_0)}\right)^2  r\ dr \right) %
\]
\begin{equation}
+ \ 2\pi r_g \sum_{m=1}^{\infty} c_{2m}^2\frac{4m(2m+1)}{4m+1}\int_{r_0}^{\infty} (r - 2r_g) %
\left( \frac{2 {\cal Q}_{2m-1}(r) + r {\cal Q}^{\prime}_{2m-1}(r)}{{\cal Q}_{2m-1}(r_0)} \right)^2 dr \ . %
\end{equation}
The potential dipole field energy $M_{\mathrm{pot}}$ can be
calculated as follows,
\begin{equation}
B_r = 2 B_0  g_r(r) \cos\theta \ ,
\end{equation}
\begin{equation}
B_{\theta} = B_0  g_{\theta}(r) \sin\theta \ ,
\end{equation}
where
\begin{equation}
g_r(r) = \frac{1}{2}\ln\left( \frac{r}{r - 2 r_g}\right) -
\frac{r_g}{r} - \frac{r_g^2}{r^2} \ ,
\end{equation}
\begin{equation}
g_{\theta}(r) = \sqrt{1 - \frac{2 r_g}{r}} \left[ \frac{2 r_g (r -
r_g)}{r (r - 2 r_g)} - \ln\left( \frac{r}{r - 2 r_g}\right)
 \right] \ .
\end{equation}
The potential dipole field energy is
\begin{equation}
M_{\mathrm{pot}} = \frac{1}{2} \int B_0^2 \left( 4 g_r^2(r)
\cos^2\theta + g_{\theta}^2(r)\sin^2\theta \right) dV \ .
\end{equation}
\section{Solutions for the Ordinary Differential Equation (\ref{ODE})}
For $m = 3, 4, 5, 6, 7,$ and 8, the functions $f_m(r)$'s are
\begin{equation}
f_3 = \frac{\lambda}{2} \ ,
\end{equation}
\begin{equation}
f_4 = \lambda \frac{2 r r_g + 2 r_g^2 + r^2\ln(r-2r_g) - r^2\ln r
}{8 r_g^3} \ ,
\end{equation}
\begin{equation}
f_5 = \lambda \frac{6 r^2 r_g + 6 r r_g^2 + 8 r_g^3 + 3 r^3
\ln(r-2r_g) - 3 r^3 \ln r }{48 r r_g^4} \ ,
\end{equation}
\begin{equation}
f_6 = \lambda \frac{6\, r^3 r_g  + 6 r^2 r_g^2
 + 8 r r_g^3  + 12 r_g^4  + 3 r^4 \ln(r-2r_g) - 3 r^4
 \ln r }{192 r^2 r_g^5} \ ,
\end{equation}
\begin{equation}
f_7 = \lambda \frac{30 r^4 r_g   + 30 r^3 {r_g}^2   + 40
r^2{r_g}^3 + 60 r{r_g}^4   + 96{r_g}^5 + 15 r^5 \ln \frac{r - 2
r_g}{r}}{2880 r^3 {r_g}^6} \ ,
\end{equation}
\begin{equation}
f_8 = \lambda \frac{30 r^5 r_g   + 30 r^4 {r_g}^2 + 40r^3{r_g}^3 +
60r^2{r_g}^4 + 96r{r_g}^5 + 160 {r_g}^6 + 15 r^6 \ln \frac{r - 2
r_g}{r}}{7680 r^4 {r_g}^7} \ ,
\end{equation}
respectively. In the calculation of the magnetic energy in the
exterior of the neutron star, the stream functions are normalized
as
\begin{equation}
\Psi = \frac{f_m(r)}{f_m(r_0)}\sin^2\theta \ .
\end{equation}



\end{document}